\documentclass[preprint,showpacs,preprintnumbers,amsmath,amssymb]{revtex4}

\usepackage{graphicx}
\usepackage{dcolumn}
\usepackage{bm}

\begin{document}


\title{Two interacting hard disks within a circular cavity: \\
towards a quantal equation of states}

\author{Takeo Kato}
\email{kato@a-phys.eng.osaka-cu.ac.jp}
\author{Naofumi Nakazono}
\author{Katsuhiro Nakamura}
\affiliation{
Department of Applied Physics, Osaka City University,
3-3-138 Sugimoto Sumiyoshi-ku, Osaka 558-8585, Japan}
\date{\today}

\begin{abstract}
We investigate a circular cavity billiard within which a pair of identical 
hard disks of smaller but finite size is confined.
Each disk shows a free motion except when bouncing elastically 
with its partner and with the boundary wall.
Despite its circular symmetry, this system is nonintegrable and 
almost chaotic because of the (short-range) interaction between the disks.
We quantize the system by incorporating the excluded volume
effect for the wavefunction.
Eigenvalues and eigenfunctions are obtained by tuning the
relative size between the disks and the billiard.
We define the volume $V$ of the cavity and
the pressure $P$, i.e., the derivative of each eigenvalue with respect to $V$.
Reflecting the fact that the energy spectra of eigenvalues
versus the disk size show a multitude of level repulsions,
$P-V$ characteristics shows the anomalous fluctuations
accompanied by many van der Waals-like peaks in each of individual
excited eigenstates taken as a quasi-equilibrium.
For each eigenstate, we calculate the expectation values of 
the square distance between two disks, and point out their relationship with
the pressure fluctuations.
\end{abstract}

\maketitle

\section{Introduction}

The study of classical and quantum billiards has a long 
history~\cite{Krylov79,Arnold68,Bunimovich80,Gaspard98,Gaspard89}.
From a viewpoint of classical dynamics, cavity billiards
are classified into two categories according to their
integrability or nonintegrability~\cite{Lichtenberg83}.
The stadium is a prototype of the nonintegrable
and chaotic billiards, whereas highly-symmetric billiards like
circle, triangle and square are integrable and regular.
The quantum-mechanical feature
of these billiards  such as level statistics is nowadays
well known~\cite{Giannoni89,Stockmann99}.

Most of the studies so far, however, are limited to the systems of a
single point particle or a single disk confined in billiards.
There exists little work on the billiards which contain
a finite number of mutually-interacting
particles~\cite{Awazu01,Ahn99,Papenbrock00}.
The interaction between particles will make the system 
nonintegrable and chaotic
even if the confining cavity is highly symmetric.
On the other hand, recent development in hightechnology has
fabricated quantum dots and optically-trapped atoms
where a finite number of interacting
particles are trapped in a small area.
Quantum mechanics of these systems constitutes a topical 
subject~\cite{Tarucha96,Kouwenhoven97}.

Considering the above circumstances, we want to understand what
kind of novel phenomena should occur
in the cavity billiards where interacting small particles are embedded.
In this paper, we consider a
pair of identical hard disks of finite size accommodated in a circular cavity.
Each disk is assumed to bounce elastically
with its partner and with the boundary wall, but otherwise showing a 
ballistic motion.
Although the system is circularly symmetric, it is nonintegrable and chaotic
thanks to the (short-range) interaction between the disks, as proved
in the following analysis of classical dynamics.

The tunable control parameter of the system is an aspect ratio of the
radii between each disk and the circular cavity.
When the aspect ratio is varied, how do the quantal eigenvalues
and eigenfunctions behave?
A multitude of level repulsions and level-spacing distributions like
Wigner distribution
will be shown to appear.
Recalling the similarity between the present system and
a nonideal gas confined within a container, the
most important feature of this system would be
a {\it quantal equation of states}, i.e.,
the volume $V$ dependence of the pressure $P$ at the cavity wall.
Under a fixed radius of each disk, tuning of the aspect ratio
corresponds to contraction
or expansion of the volume $V$
of the circular cavity.
The pressure $P$ at the
wall boundary is obtained from the derivative of
parameter-dependent eigenvalues with respect to $V$.
In the classical nonideal gas theory, $P$ does not decrease monotonically
as $V$ is increased, and is accompanied
by the van der Waals peak. This peak appears due to
the competition between the
finite size of each molecule and the attractive interaction
between molecules.
In the present system, we find the competition between
the finite size of each disk and
the quantum-mechanical correlation due to symmetrization
of two-particle wavefunctions.
Therefore $P-V$ characteristics in the present system is 
expected to exhibit novel
fluctuations unseen in ordinary quantum billiards
that contain only a single particle or disk.
This is an advantage of the interacting disk systems over
the ordinary quantum billiards.
By using two-particle wave functions, we shall evaluate
expectation values for the distance between disks, which
will elucidate a physical mechanism
for fluctuations in the quantal equation of states.

The organization of the paper is as follows:
In Section~\ref{model}, we shall introduce a model system
and analyze the underlying classical dynamics.
We shall compute the maximum Lyapunov exponent
with use of a modern technique 
by Gaspard and Beijeren~\cite{Gaspard02}.
In Section~\ref{technique} we define the two-particle basis wavefunctions
appropriate to quantize this complicated system and describe
some technical details
to obtain a standard form of the eigenvalue equation.
The result for eigenvalues is given for various
aspect ratios in Section~\ref{eigenvalue}.
Here we shall also analyze $P-V$ characteristics, which
will be the central theme of our study.
Section~\ref{eigenfunction} is devoted to investigation of
expectation values for the distance between disks, where
the single-particle density and two-particle correlation functions are 
exploited. The final Section is concerned with summary and discussions.

\section{Model system and classical dynamics}\label{model}

\begin{figure}[hbp]
\begin{center}
\includegraphics[width=60mm]{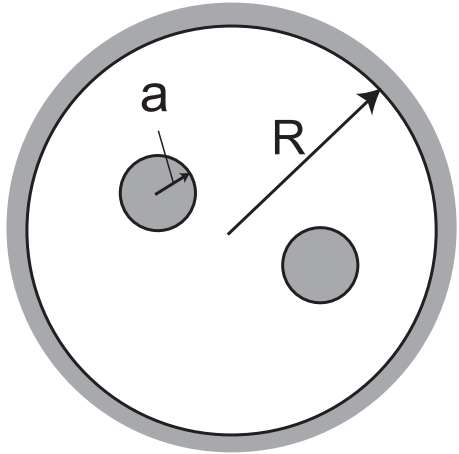}
\end{center}
\caption{Model system.}
\label{fig:1}
\end{figure}

\begin{figure}[tbp]
\begin{center}
\includegraphics[width=8cm]{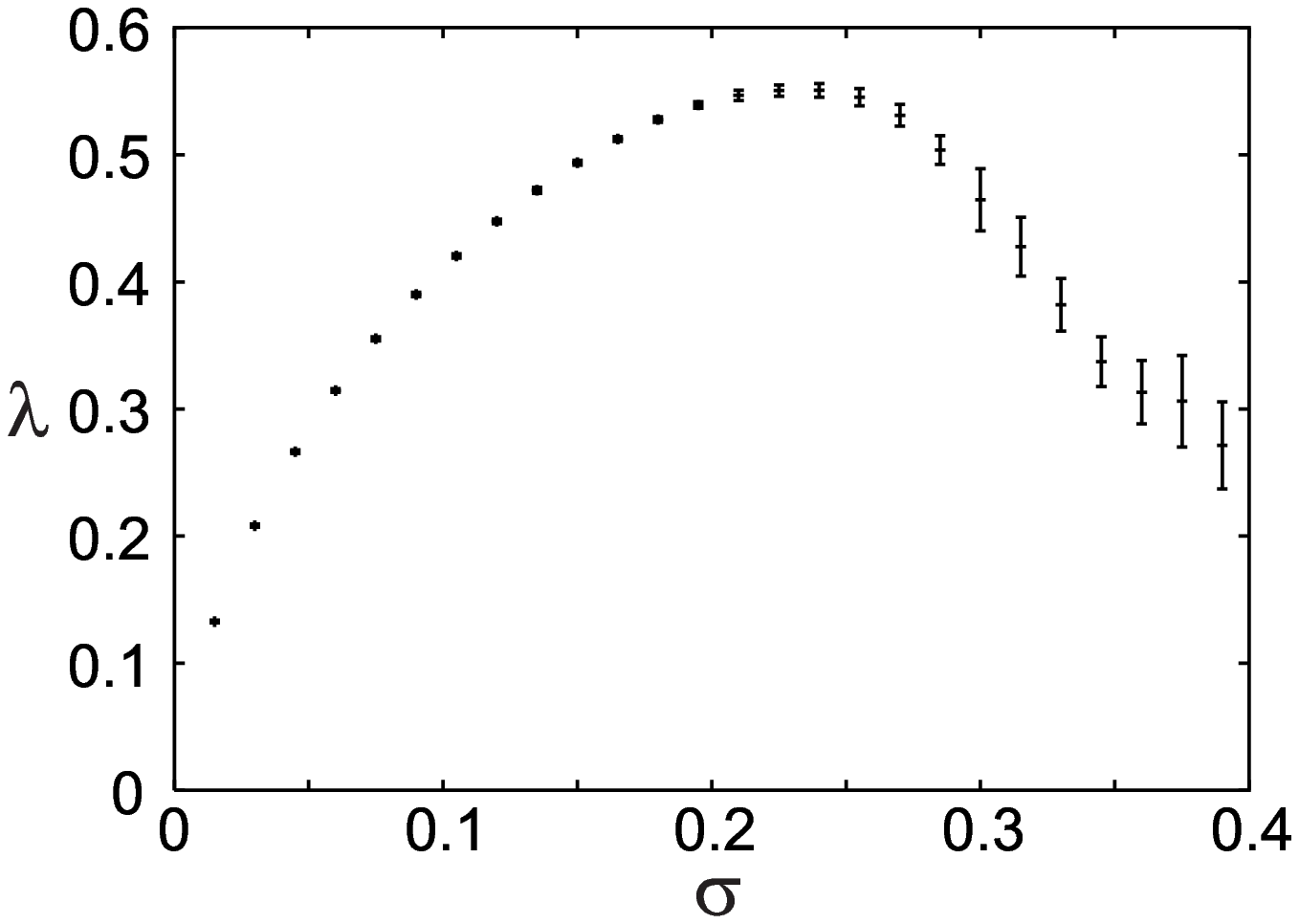}
\end{center}
\caption{The ratio $\sigma = a/R$ vs. the maximum Lyapunov exponent
$\lambda$ calculated by the numerical simulation.
The total $L_z$ is taken as zero. The error bar means the standard 
deviation of obtained $\lambda$'s.}
\label{fig:Lyapunov}
\end{figure}

Figure~\ref{fig:1} illustrates our model system consisting of
two identical hard disks moving within a circle billiard.
Each disk shows a ballistic motion except when bouncing 
with its partner and with the 
billiard boundary.  Let $R$ and $a(<R/2)$ be the radii for 
the circle billiard and each disk, respectively.
Here the internal self-rotation (spinning motion) of each disk are ignored.
The single-disk motion in the circle billiard is integrable since
the number of degrees of freedom (two) accords with 
that of constants of motion, i.e., energy and
angular momentum. However, the two-disk system proposed here
has the degree of freedom (four) which is larger than
the number of constants of motion, i.e., the total energy, 
angular momentum. Therefore the system becomes nonintegrable, and
may be chaotic.

The chaotic dynamics is characterized by the positive maximum Lyapunov
exponent $\lambda$. This can be obtained by numerical simulation for a
given initial condition by using linearized equations for deviations
of an orbit. Although the degrees of freedom is four in the present system,
the calculation of $\lambda$ is greatly simplified because of the 
nature of billiard motion~\cite{Gaspard95,Gaspard02}. 
For the evaluation of $\lambda$,
each orbit is calculated over $N_{\rm col}=10^6$ collisions. Generally,
$\lambda$ depends on the initial conditions. We randomly choose 
$N_{\rm init}=10^5$ initial conditions satisfying $L_z = 0$.
The average and standard deviation of obtained $\lambda$'s are
shown in Fig.~\ref{fig:Lyapunov}. For $0<\sigma=a/R<0.4$, the positive
maximum Lyapunov exponent can be observed. The standard deviation
systematically decreases below $\sigma \sim 0.25$ as $N_{\rm col}$ increases.
The convergence, however, is not observed above $\sigma\sim 0.25$ as 
shown by large standard deviations; This indicates that 
the ergodicity is not guaranteed in this region. 

Through the numerical simulation, we found that a small amount of 
initial conditions leads to $\lambda = 0$ indicating the existence of
tori in the phase space in the full region of $\sigma$. 
The ratio of tori to the whole phase space is, however, extremely small,
and is estimated typically as smaller than of $10^{-3}$.
Although the present system should be said exactly to be a mixed system,
the chaotic sea occupies almost whole part of the phase space, and
one can proceed to investigate a quantum analogue of chaos in this system.

\section{Methodology of quantization}\label{technique}
 
To make variables dimensionless, 
we choose the scaling for coordinates and energy as
\begin{eqnarray}
{\bf r}^{\prime} = \frac{{\bf r}}{R-a}, &\quad &
a^{\prime} =\frac{a}{R-a}, \nonumber \\
R^{\prime} =\frac{R}{R-a}, &\quad & 
E^{\prime} =\frac{E}{\frac{{\hbar}^{2}}{2m(R-a)^2}}.
\label{eqn:2}
\end{eqnarray}
Note that by this scaling the ratio $\sigma=a/R=
a^{\prime}/R^{\prime}$ is not changed.
By the above scaling, the admissible range
for $r^{\prime}=|{\bf r}^{\prime}|$ becomes 
$0 \le r^{\prime} \le 1$. Hamiltonian is then given in a 
dimensionless form as
\begin{equation}
H=-\sum_{i=1,2}
\left(\frac{\partial^2}{\partial r^{\prime 2}_i}+
\frac{1}{r^{\prime}_i} \frac{\partial}{\partial r^{\prime}_i}+
\frac{1}{r^{\prime 2}_i}\frac{\partial^2}{\partial \theta^{\prime 2}_{i}}
\right)
+\sum_{i=1,2}U({\bf r}^{\prime}_i) + V(|{\bf r}^{\prime}_1 
- {\bf r}^{\prime}_2|),
\label{eqn:3}
\end{equation}
where
\begin{equation}
U({\bf r}'_i)=
\left\{
\begin{array}{cl}
\infty & (|{\bf r}'_i| \ge 1)\\
0 & (|{\bf r}'_i| <1)
\end{array}
\right. ,
\label{eqn:4}
\end{equation}
and
\begin{equation}
V(\xi)=
\left\{
\begin{array}{cl}
0 & (\xi \ge 2a')\\
\infty & (\xi<2a')
\end{array}
\right. .
\label{eqn:5}
\end{equation}
The second and third terms in (\ref{eqn:3}) represent confining potential 
by the hard wall of the circle billiard and
inter-disk interaction due to their hard cores, respectively.
In the case of a single disk inside the circle billiard,
the appropriate wavefunction satisfying the boundary condition
is given by Bessel function of integer order for the radial part 
multiplied by the angular function as
\begin{equation}
J_{k}(\lambda_{kn}r')e^{ik\theta'},
\label{eqn:6}
\end{equation}
where $k$ and $n$ are integers and $J_{k}(\lambda_{kn}r')$ vanishes at $r'=1$, 
namely at the boundary wall.
By choosing the wavefunction (\ref{eqn:6}) for each of the disks, Dirichlet
boundary condition
represented by the potential $U({\bf r}'_i)$ is automatically satisfied. 
On the other hand, the basis functions for the two-disk system are given 
by a product of (\ref{eqn:6}) and take
\begin{equation}
\Phi_{\alpha}=J_{k_{1}}(\lambda_{k_{1}n_{1}}r'_1)J_{k_{2}}
(\lambda_{k_{2}n_{2}}r'_2)
 e^{ik_{1}\theta'_{1}}e^{ik_{2}\theta'_{2}}f(X),
\label{eqn:7}
\end{equation}
which are as yet not orthonormal.  The multiplicative factor
$f(X)$, which represents the excluded volume effect caused by the
hard cores of the disks, is defined by
\begin{eqnarray}
f(X) =
\left\{
\begin{array}{cl}
\ 1-e^{-\beta X} &  (X > 0)\\
0 & (X \le 0)
\end{array}
\right. .
\label{eqn:8}
\end{eqnarray}
Here $X \equiv r^{\prime 2}_{1}+r^{\prime 2}_{2}-2r^{\prime}_{1}
r^{\prime}_{2}\cos \phi-l_{0}^{2}$ with
$\phi=\theta^{\prime}_2-\theta^{\prime}_1$ and $l_0=2a'$.
$X$ stands for a void between two hard disks,
$\phi$ is a relative angle, and
$l_0$ is the admissible minimum distance 
between the centers of two hard disks, 
at which they touch each other.
$\beta$ is arbitrary positive real number, and may be called as
the fictitious inverse temperature. We will 
discuss about $\beta$ at the end of this Section.
$\Phi_{\alpha}$ vanishes at $X=0$,  
satisfying another boundary condition that the wavefunction 
for two disks should vanish when they touch each other. In other words,
the inclusion of the factor $f(X)$ is approximately identical to incorporating 
the effect of the short-range repulsive interaction
$V(|{\bf r}^{\prime}_1 - {\bf r}^{\prime}_2|)$ in (\ref{eqn:3}). 
Finally we symmetrize the wavefunction noting the
indistinguishability of two disks.
We here choose a symmetrized wave function by assuming the invariance 
of wavefunctions against the exchange
of the disk coordinates. This choice is appropriate when the disks are
Boson or Fermion forming a spin-singlet state.
Therefore we use the basis function defined by
\begin{equation}
\Psi_{\alpha}=(\Phi_{\alpha}(1,2)+\Phi_{\alpha}(2,1))/\sqrt{2}.
\label{eqn:9}
\end{equation}

With use of (\ref{eqn:9}) we shall proceed to construct the energy matrices.
First, by operating the Hamiltonian (\ref{eqn:3}) on (\ref{eqn:9}), we find
\begin{eqnarray}
H\Psi_{\alpha}
&=&({\lambda_{k_{1}n_{1}}}^{2} + {\lambda_{k_{2}n_{2}}}^{2})\Psi_{\alpha}
\nonumber \\
&+& \frac{1}{\sqrt{2}} \Biggl[ \, \biggl\{ -8\beta J_{k_{1}}J_{k_{2}}
e^{ik_{1}\theta^{\prime}_{1}}e^{ik_{2}\theta^{\prime}_{2}} e^{-\beta X}
(1-\beta ({r^{\prime}_{1}}^{2}+{r^{\prime}_{2}}^{2}-
2r^{\prime}_{1}r^{\prime}_{2}\cos \phi))
\nonumber \biggr. \Biggr.
\\
& & -4\beta e^{ik_{1}\theta^{\prime}_{1}}e^{ik_{2}\theta^{\prime}_{2}}
e^{-\beta X} \left( J_{k_{2}}
\frac{\partial J_{k_{1}}}{\partial r^{\prime}_{1}}
(r^{\prime}_{1}-r^{\prime}_{2}\cos \phi)
+J_{k_{1}}\frac{\partial J_{k_{2}}}
{\partial r^{\prime}_{2}}(r^{\prime}_{2}-r^{\prime}_{1}
\cos \phi) \right)
\nonumber \\
& & \biggl. + 4i\beta J_{k_{1}}J_{k_{2}}e^{ik_{1}\theta^{\prime}_{1}}
e^{ik_{2}\theta^{\prime}_{2}}e^{-\beta X}\sin \phi
\left(\frac{r^{\prime}_{2}}{r^{\prime}_{1}}k_{1}-
\frac{r^{\prime}_{1}}{r^{\prime}_{2}}k_{2}\right)
\biggr\}  \nonumber
\\
& &+ \Biggl. \biggl\{ 1 \leftrightarrow 2\ \biggr\} \Biggr].
\label{eqn:10}
\end{eqnarray}
Second, multiplying (\ref{eqn:10}) by "$bra$" that corresponds to 
the complex conjugate of (\ref{eqn:9}), we
integrate each term over coordinates.
The integration about angles $\theta^{\prime}_1$, $\theta^{\prime}_2$ can be 
performed analytically.
But we have to carry out the numerical integration
about the radial coordinates
$r^{\prime}_1$, $r^{\prime}_2$.
Third, noting that the total angular momentum $L_z$ is a good quantum number, 
we diagonalize the energy matrices for each of the fixed value $L_z$.
That is $ k_{1}+k_{2}=k_{1}^{\prime}+k_{2}^{\prime}
=L_z $ with $L_z=0,\pm 1, \pm 2$, and so on.

Finally, before proceeding to diagonalization of each energy matrix,
it must be regularized
because we are using non-orthogonal basis functions. 
This is performed as follows:
In the eigenvalue problem $H\Psi=E\Psi$ under consideration,
we substitute the expansion $\Psi =\sum_{\alpha}C_{\alpha}\Psi_{\alpha}$
with $\Psi_{\alpha}$ the non-orthogonal basis functions.
Then one reaches
\begin{eqnarray}
\sum_{\alpha}H_{\alpha^{\prime}\alpha}C_{\alpha}
=\sum_{\alpha}EN_{\alpha^{\prime}\alpha}
C_{\alpha},
\label{eqn:11}
\end{eqnarray}
with 
$H_{\alpha^{\prime}\alpha}=
\langle \Psi_{\alpha^{\prime}}|H|\Psi_{\alpha}\rangle$
and
$\langle \Psi_{\alpha^{\prime}}|\Psi_{\alpha}\rangle
=N_{\alpha^{\prime}\alpha}\neq \delta_{\alpha^{\prime}\alpha}$.

With use of the diagonalized norm-kernel $N_d$ obtained by
\begin{equation}
U^{-1}NU \rightarrow N_{d},
\label{eqn:12}
\end{equation}
we introduce the regularized energy matrix 
\begin{equation}
\tilde{H}=N_{d}^{-1/2} U^{-1} H U N_{d}^{-1/2}.
\label{eqn:13}
\end{equation}
As a consequence, we arrive at a standard form of the eigenvalue equation
\begin{equation}
\tilde{H}{\bf g}=E{\bf g},
\label{eqn:14}
\end{equation}
where ${\bf g} = N_d^{1/2} U^{-1} {\bf C}$.
This equation will be solved in the next Section.

Before closing this Section, we should comment on
the fictitious inverse temperature $\beta$ involved in the
excluded-volume effect factor $f(X)$ in (\ref{eqn:8}).
We find the calculated eigenvalues are stable against variation of $\beta$
so long as it falls in $0.5<\beta<3.0$ so that we shall fix to $\beta=1.5$.

\section{Energy spectra and pressure}\label{eigenvalue}

\subsection{Eigenvalues and level-spacing distribution}
\label{level spacing}

\begin{figure}
\begin{center}
\includegraphics[width=120mm]{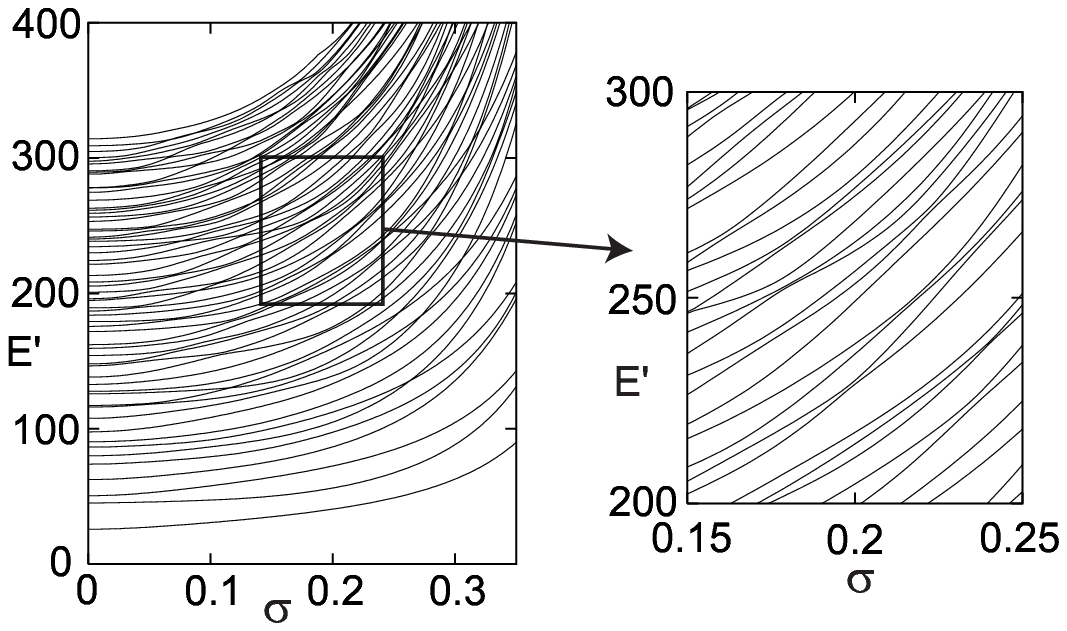}
\caption{Energy levels versus ratio of radii ($\sigma$) between each disk
and a circle billiard in the manifold $L_z=1$.
Inset is a partial magnification.}
\end{center}
\label{fig:spectrum}
\end{figure}

We prescribe $\sigma=a/R$ with $0\le\sigma \le 0.5$
as a tunable parameter controlling the size  of the disk. 
Under a fixed angular momentum $L_z$,
we diagonalize the energy matrix in $10^3 \times 10^3 $ dimensions 
by varying $\sigma$.
The lower half of $10^3$ eigenvalues, which have a precision of 4 digits, 
is used for our study below.
In the manifold $L_z=1$, the energy spectrum against $\sigma$ 
is given in Fig.~3.
The upward shifts of energies as a whole with increasing $\sigma$ reflect
the confining of disks in a rapidly-decreasing effective area 
inside the wall boundary.
This spectrum together with its partial magnification
shows a multitude of avoided level crossings (level repulsions).
Except for the lower energy region, the avoided crossings are 
widely seen in the full range of $0 < \sigma < 0.5$.

\begin{figure}
\begin{center}
\includegraphics[width=130mm]{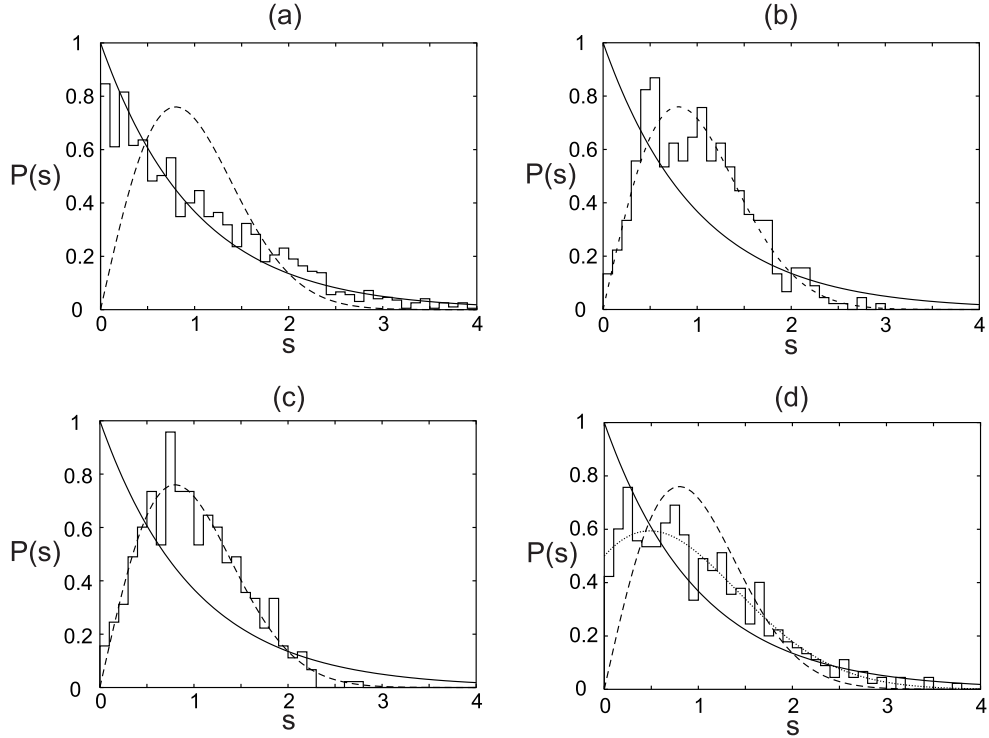}
\caption{Histogram of level-space distribution:
(a)Noninteracting disks with $\sigma=0.2$ and $L_z=1$;
(b)Interacting disks with $\sigma=+0$ and $L_z=1$;
(c)Interacting disks with $\sigma=0.2$ and $L_z=1$;
(d)Interacting disks with $\sigma=0.2$ and $L_z=0$.
The solid, dashed and dotted curves show the Poisson distribution, 
Wigner distribution and the mixture of two independent Wigner distributions,
respectively.}
\end{center}
\label{fig:distribution}
\end{figure}

We proceed to investigate the level-spacing distributions, 
where the usual unfolding 
procedure is performed~\cite{Haake00,Mehta91}.
Figures~4~(a)-(c) are the level-spacing distributions
$P(s)$ for typical values of $\sigma$ with $L_z=1$, 
obtained from intermediate $450$ levels 
by suppressing both the lowest $50$ levels and the upper half 
of $10^3$ levels. Figure~4~(a) is 
the result for the noninteracting
two-disk system within the circle billiard, which approximately
provides Poisson distribution $P(s)=\exp(-s)$. Small deviation 
from the Poisson distribution is however observed; We infer that 
this deviation comes from the particular properties of the zero points
of the Bessel function which tends to be periodic in the asymptotic limit. 
In the case of interacting disks under consideration, the Wigner distribution 
$P(s)=\frac{\pi}{2}\exp (-\pi s^2)$ can be seen in the full range of
$0\le\sigma \le 0.5$. This distribution is known as a quantal signature 
of chaos.
It is interesting that Wigner distribution is obtained even in the 
point-disk limit, $\sigma=+0$. This is due to the fact that, in this limit,
the present system has a short-range interaction of delta-function type,
and is distinct from a noninteracting two-disk system.
The stability of level statistics against the variation of $\sigma$
is very convenient in our study below on the pressure.

Before proceeding further, however, we should note an atypical
level statistics in the manifold with $L_z=0$.
Figure~4~(d), with a finite
weight at the origin, is neither Wigner nor Poisson
distribution. This distribution can be fitted by the mixture
of two independent Wigner distributions (shown by the dotted 
curve in Fig.~4~(d)).
We expect that the origin of this feature comes from the existence of
a kind of the time-reversal symmetry. We can define a operator $T$ commuting 
with the Hamiltonian as
\[ T \Psi_{k,n_1,-k,n_2} = \Psi_{k,n_2,-k,n_1}, \]
where $(k,n)$ denotes the one-particle state, and $\Psi_{k,n_1,-k,n_2}$
is a (symmetrized) two-particle basis function whose total $L_z$ equals to
zero. The eigenstates can be divided into
two kinds by the eigenvalue of $T$, and the level spacing
distribution becomes the mixture of two independent Wigner distributions.

\subsection{Pressure versus volume}

\begin{figure}
\begin{center}
\includegraphics[width=80mm]{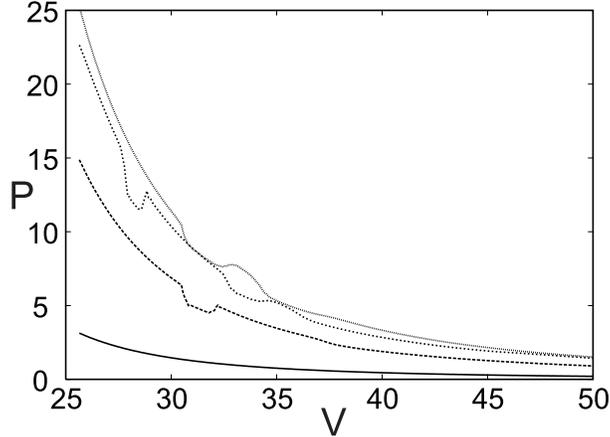}
\caption{Pressure versus cavity volume. From the bottom to the top,
the level number is taken as 1, 19, 38 and 45.}
\end{center}
\label{fig:PV1}
\end{figure}

When a disk hit the cavity wall elastically, 
its momentum is reversed, which is compensated by the impulse on the wall.
This is the origin of pressure on the wall.
In the classical ideal gas theory, the pressure $P$ is inversely proportional 
to the volume $V$. In the nonideal gas, on the other hand, 
there exists a competition between two kind of interactions, namely,
the short-range repulsion due to hard cores of constituent molecules
and the long-range attractive interaction between them. This competition 
leads to a peak called van der Waals peak in $P-V$ characteristics.
In the quantum two-disk system under consideration, we have
both the hard-core repulsion and the strong quantum 
correlation due to symmetrization of two-disk wavefunctions, 
and naturally can expect van der Waals-like peaks in $P-V$ characteristics.
The pressure $P$ in each excited state taken as an equilibrium
is obtained by means of the $\sigma$-dependent eigenvalues.
We choose $a=1$, and define the cavity volume as $V = \pi R^2$.
Then, the pressure is calculated as
\begin{eqnarray}
P=-\frac{d E_l}{d V}=-\frac{d E_l}{d \sigma}
\frac{d \sigma}{d R} \frac{dR}{dV} 
=\frac{1}{2\pi R^3}\frac{d E_l}{d \sigma},
\label{pressure}
\end{eqnarray}
where $E_{l}$ stands for the energy for level $l$.
Figure 5 shows $P-V$ characteristics for several eigenstates.
From the definition (\ref{pressure}), the van der Waals-like peaks and 
pressure fluctuations are attributed to level repulsions.
In the low-lying states, there is no van der Waals-like peak, reflecting
the absence of level repulsions. In the high-lying states, however,
the number of peaks are increased, in contrast to
the classical nonideal gas theory that accommodates only a single peak. 

A comment should be made here: In the case of a single disk 
in chaotic billiards like
a stadium, one might also see the avoided crossing (AC) and level fluctuations
as the system's parameter
is varied. However, it is difficult to find its analogy to 
the nonideal gas because of the absence of
inter-disk interactions. Furthermore,
most of chaotic billiards have distorted wall boundary, for which
there exists no global pressure as defined in (\ref{pressure}) 
in the case of circle billiard.
Thus $P-V$ characteristics and the analogy to the nonideal gas 
should be meaningful only
in the case of interacting particles (disks) confined in 
the circle billiard.

\section{Eigenfunction}
\label{eigenfunction}

\begin{figure}
\begin{center}
\includegraphics[width=80mm]{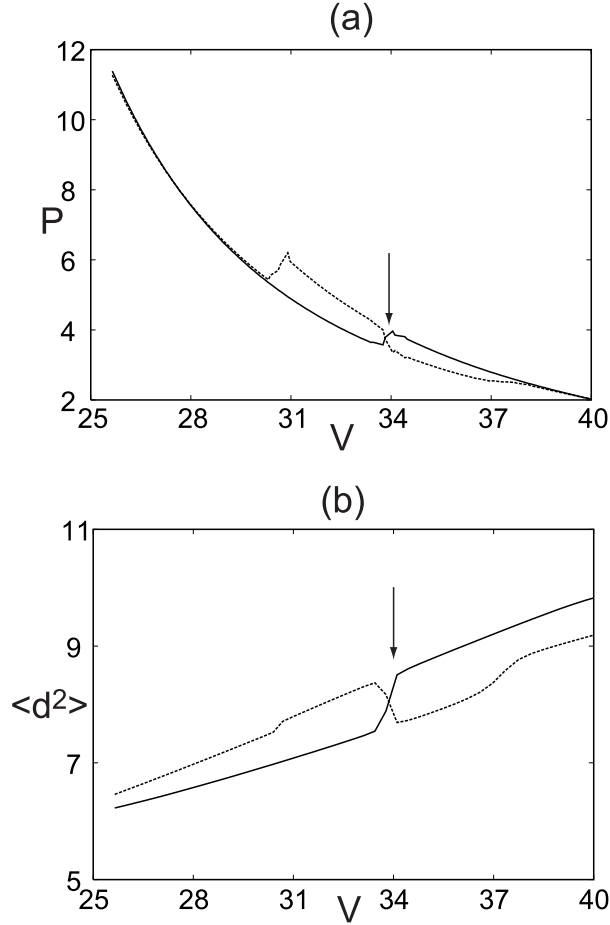}
\end{center}
\caption{(a) The pressure versus volume for a level $17$ (the solid curve) 
and $18$ (the dashed curve). (b) Corresponding average of the distance 
between two disks. The arrow shows the level crossing between these two
levels.}
\label{fig:PV2}
\end{figure}

The fluctuations of the pressure mentioned above is a 
promising manifestation of quantum chaos in interacting 
two-disk systems in the billiard. In this section,
we discuss this pressure fluctuation in terms of the wave function.

Although the two disks originally have repulsive correlation due 
to hard-core interaction, quantum effective exchange interaction
gives  an additional strong quantum correlation between disks. 
The averaged distance between disks is determined
by the competition between repulsive hard-core interaction and
quantum correlation depending on the control parameter $\sigma$. 
We now wish to make clear whether this competition is responsible for
the van der Waals-like peak in quantum systems.
For this purpose, we investigate the expectation values of the
square distance 
\[ \langle d^2 \rangle = \int 
r_1 {\rm d}r_1 r_2 {\rm d}r_2 {\rm d}\theta_1 {\rm d}\theta_2 \
d^2 |\psi(r_1,r_2,\theta_1,\theta_2)|^2 \]
for each level, where $d = |{\bf r}^{\prime}_1 - {\bf r}^{\prime}_2|$.

We concentrate on typical two neighboring levels employed in Fig.~6~(a), 
which bear sharp avoided crossings (AC) at the point indicated by
the arrow. The avoided level crossing leads to the van-der-Waals-like
peak (or dip). In order to study of the origin of this peak, 
we show the average of the distance between the disks in Fig~6~(b).
Clearly, the sharp changes in the pressure curve corresponds to the 
change in the averaged distance; When the averaged distance 
increase(decrease) sharply, the pressure increases(decreases) similarly.
This result is intuitively reasonable; If the averaged distance 
increases, the repulsive correlation between disks works strongly, and the
existence probability of the disks near the circular wall will increase. 
Hence, the repulsive correlation contributes to the upward jump
in the pressure curve.

\section{Summary and discussions}\label{summary}

Quantum mechanics of a pair of identical hard disks confined
in a circular  billiard 
is investigated. Although the system is circularly symmetric, 
the short-range interaction between the disks turns out making it
nonintegrable and almost chaotic.
With use of the two-particle basis wavefunctions that 
incorporate the excluded volume effect,
eigenvalues and eigenfunctions are obtained for various ratios of 
the radii between each disk and the circle billiard.
The energy spectra of eigenvalues versus the disk radius
show a multitude of level repulsions.
We find the Wigner level spacing distribution in the full 
range of the disk radius $0<a<R/2$.
The Wigner distribution in the vicinity of $a=+0$ should be ascribed to
the strong inter-disk repulsion prevailing even at $a=+0$.

The most important feature of this system is a {\it quantal equation 
of states}, i.e., the volume $V$ dependence of the pressure $P$.
The tuning of the aspect ratio corresponds to contraction 
or expansion of the cavity volume $V$. This tuning also changes
the pressure $P$ felt by the
wall boundary. $P-V$ characteristics shows novel
fluctuations coming from quantum level repulsions.
The van der Waals-like peaks in this fluctuations 
are attributed to the competition 
between the finite size of two disks 
(responsible for the short-range repulsion) and
the quantum-mechanical attractive correlation between them.
The expectation values of the square distance between the disks is 
found to elucidate a mechanism for the pressure fluctuations.

\end{document}